\documentclass[manuscript,natbib=false]{acmart}
\AtBeginDocument{%
  \providecommand\BibTeX{{%
    \normalfont B\kern-0.5em{\scshape i\kern-0.25em b}\kern-0.8em\TeX}}}

\usepackage[style=acmnumeric]{biblatex}
\setcopyright{acmcopyright}
\copyrightyear{2022}
\acmYear{2022}

\acmConference[RecSys '22]{16th ACM Conference on Recommender Systems}{September 18--23, 2022}{Seattle, WA, USA}
\usepackage[acronym]{glossaries}    
\usepackage{subcaption}

\makeglossaries
\newacronym{nft}{NFT}{Non-fungible Token}
\newacronym{ml}{ML}{Machine Learning}
\newacronym{dl}{DL}{Deep learning}
\newacronym{ai}{AI}{Artificial Intelligence}
\newacronym{nlp}{NLP}{Natural Language Processing}
\newacronym{erc}{ERC}{Ethereum Request for Comments}
\newacronym{lstm}{LSTM}{Long short-term memory}
\newacronym{api}{API}{Application Programming Interface}
\newacronym{recsys}{RecSys}{Recommendation System}

\begin{document}

\title[NFT-Trends-RecSys]{Exploration of the possibility of infusing Social Media Trends into generating NFT Recommendations}

\author{Dinuka Ravijaya Piyadigama}
\email{drpiyadigama@gmail.com}
\orcid{0000-0002-8123-0099}
\affiliation{%
  \institution{University of Westminster}
  \streetaddress{309 Regent St.}
  \city{London}
  \country{UK}
}

\author{Guhanathan Poravi}
\affiliation{%
  \institution{Informatics Institute of Technology}
  \streetaddress{57 Ramakrishna Rd}
  \city{Colombo 06}
  \country{Sri Lanka}}
\email{guhanathan.p@iit.ac.lk}


\begin{abstract}
Recommendations Systems have been identified to be one of the integral elements of driving sales in e-commerce sites. The utilization of opinion mining data extracted from trends has been attempted to improve the recommendations that can be provided by baseline methods in this research when user-click data is lacking or is difficult to be collected due to privacy concerns.

Utilizing social trends to influence the recommendations generated for a set of unique items has been explored with the use of a suggested scoring mechanism. Embracing concepts from decentralized networks that are expected to change how users interact via the internet over the next couple of decades, the suggested Recommendations System attempts to make use of multiple sources of information, applying coherent information retrieval techniques to extract probable trending items.

The proposed Recommendations Architecture in the research presents a method to integrate social trends with recommendations to produce promising outputs.
\end{abstract}




\maketitle

\section{Introduction}

\gls{nft}s allow people to trace the origin of digital items with the help of Blockchain technology. Since the introduction of crypto, \gls{nft}s have stood out to be the most widely accepted application of Blockchain technology.

One \gls{nft} is expected to be unique from another. As these items are unique from each other, as expressed by the name itself, they are \textit{not fungible}. They cannot be replaced like crypto, which is fungible.

Due to several restraints that are presented with the nature of \gls{nft}s \& the overwhelming amount of data that needs to be analyzed, it is difficult to find NFTs of comparable value that are trending among the community, timely and relevant to each user’s identified interests.

Recommendations Systems have been identified to be one of the integral elements of driving sales in e-commerce sites. They have been driving engagement and consumption of content as well as items on almost every corner of the internet over the last decade. 30\% of Amazon's revenue is said to come from the items recommended to users \cite{naumovDeepLearningRecommendation2019}. 60\% of watch time on Youtube and 75\% on Netflix were also reported to have come about as a result of recommendations \cite{RecommendationsWhatWhy, vanderbiltScienceNetflixAlgorithms}.

In the month of June of 2021, OpenSea which is poised to be the Amazon of \gls{nft}s facilitated sales of \$150 million and was valuated at \$1.5 Billion \cite{hackettThisCryptoMarketplace2021, dfinzerAnnouncingOur100M2021, chevetBlockchainTechnologyNonFungible2018}. Therefore, it is clear that Recommendation Systems could help bolster sales on such \gls{nft}-marketplaces, bringing in more revenue to businesses, and creators while helping users explore trending \& relevant items.

\section{Review of Related Work}

\subsection{Identified Challenges \& Requirements to build a Social Trends aware Recommendations System for NFTs}
When considering possible options to explore the possibility of recommending \gls{nft}s, it appeared that there hadn't been much past work that could relate to this specific purpose.

\begin{quote} 
\centering 
\emph{"Crypto has a founding tradition of emphasizing freedom and privacy. Maybe because of this prevailing cultural trend, the NFT space does not have many recommender systems."} 
\\
\raggedleft
\cite{WhatAreYou2020}
\end{quote}

Being loosely related to the crypto market and community \cite{dowlingNonfungibleTokenPricing2021} it is understood that the early adopters \& pioneers in this space are concerned about their privacy. 


Since \gls{nft}s have a distant relationship with crypto assets, it is expected to be of help to understand how crypto assets are evaluated when opted for selection to comprehend how \gls{nft} assets could be evaluated. A study which was done related to a modelling framework that exposes this area of research \cite{bartolucciModelOptimalSelection2020} assumes that two main features, namely security and stability can be used to determine the user's desire to own a specific crypto asset. Crypto-related assets have a tendency to change with time, social acceptance and trends. Therefore, it is important to consider these factors when building a crypto-related Recommendations System.

\subsection{User Opinion \& Sentiment Aware Recommendation Systems}

\begin{quote} 
\centering 
\emph{"Catching opinions from social media could be a cheap, fast and effective way to collect feedbacks from users"} 
\\
\raggedleft
\cite{zhangOpinionMiningSentiment2018}
\end{quote}

When the above fact is looked at in a more generalized form, it is clear that exploiting user trends that build-up of opinions from social media can lead to better quality recommendations, while \cite{huReviewerCredibilitySentiment2020} expressing how sentiment analysis of user reviews can be used to point to the direction of personalized recommendations.\\

The utilization of opinion mining data extracted from trends to improve the recommendations that can be provided by baseline methods was expected to address the restraints of recommending items presented by the very nature of \gls{nft}s.

There have been many attempts to expand the capabilities of Recommendations by making use of public opinion. Collaborative Filtering was one approach to achieve that and it has been the standard baseline technique for Recommendations for over a decade \cite{lindenAmazonComRecommendations2003, smithTwoDecadesRecommender2017}. But, it can't be taken as the only recommendations model in this use-case because, by the time one \gls{nft} is viewed many times by other users, it may already be too late for another user to purchase that item for a profitable cost as the value would've sky-rocketed due to high demand over a long period of time.

User data on social media has been integrated into \gls{ml} Hybrid Recommendation Architectures in several ways to produce user-opinion \& social context-aware Recommendations. 

One of these methods was to apply opinion mining \& sentiment analysis on users' reviews to create a preference profile and create collaborative filtering like recommendations \cite{chengHybridRecommenderSystem2020a}. While this method is effective in dealing with insufficient data, the issue related to the required use case is that users still have to place reviews on previous movies to create a preference profile, which would add to the privacy concern mentioned before. A similar \gls{dl} model was attempted to generate possible user ratings, again based on user comments \cite{chenUserRatingClassification2019}, letting the same issue prevail.

A hybrid approach of combining content-based recommendation, user-to-user collaborative filtering and personalized recommendation techniques has been attempted, to finally show the sentiment analysis polarity of the recommended item based on a user's tweets \cite{ayushiCrossDomainRecommendationModel2018}. While this model does a good job in addressing the limitation of single domain analysis such as data sparsity \& cold start problem, it doesn't consider the sentiment for the particular recommendation. It is only calculated and made visible to the user. The user is required to make the decision of selecting if the item is worth. It also doesn't make use of trends that are happening on social media.

\subsection{Research Motivation}
In recent research done by \textbf{Amazon} \cite{larryHistoryAmazonRecommendation2019} it is understood that when a timeline is considered for recommendations, an \textbf{\emph{Autoencoder} Deep Learning model} is capable of Recommending the best possible combination of movies to users. Chronologically sorted movie-viewing data managed to outperform item-to-item collaborative filtering applied with the bestseller list.
This was done by getting the model to recommend at least 2 recently released movies. The idea behind this was that a user is more likely to watch a recently released movie rather than a very popular and highly rated old movie.

The method followed to recommend items to the user in this case motivated the authors to pursue to attempt to infuse social media trends into Recommendations. The reason for this was that trends on social media will give an idea of the things, people that are very popular at that moment in time. A person would be more interested in getting a recommendation with whatever that is related to a popular topic rather than an old item that was very popular back for a while and highly rated.

This can be applied even to an e-commerce setting. But, especially in the case of \gls{nft}s, this opens up another way to get items that may not otherwise surface in front of users' eyes. It would also keep updating over time, as trending topics change.
Therefore, new trends may open up new valuable, relevant items rather than old items that are already high in demand and owned by owners who may not wish to sell.

% ...

\section{Proposed System Architecture Design}

In the attempt to find a method to recommend items based on people's aggregated opinions in the form of trending topics on social platforms, without having to track user clicks and online behaviour in a way that can expose individuals, the author decided to move in the direction of searching possible methods to integrate social media trends data that is sourced from external sources for recommendations.
Due to the privacy concern mentioned earlier, the author's goal was to build the Recommendation System in a way that potential buyers' privacy isn't threatened by the collection of user click data.

\subsection{System Process Flowchart}

\begin{figure}[h]
\centering
\includegraphics[width=\linewidth]{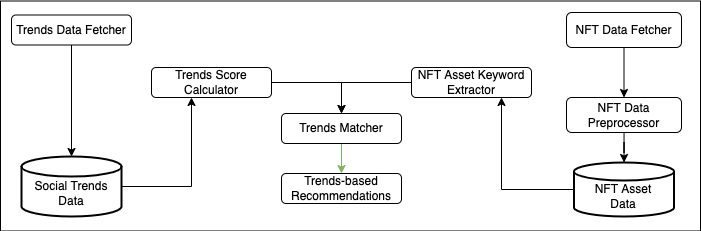}
\caption{System Process Flowchart \textit{(self-composed)}}
\label{fig:trends-recsys-system-process-flowchart}
\end{figure}

The flow of data in the system with the various processing steps these data flows through has been represented in Figure \ref{fig:trends-recsys-system-process-flowchart}.

\subsection{Algorithm Design}

The following equation was designed to calculate the impact of a trend.
\begin{equation}
i_{t} = \frac{t_{vt,c}}{Med(T_{vt})}
\end{equation}

The volume of a trend is divided by the median volume here to get a relative impact of a trend. When the trend score is kept as low as possible, when applied to the next total trend-score calculation equation, the score will drop down to a low value faster.

For trends that don't have a measurable volume, $t_{vt,c}$ can be taken as $\left(T_{vt}{min} - 1\right)$ to give it the lowest possible value, or as $Med(T_{vt})$ to omit the impact score all-together.\\

The following equation was designed to calculate the total trends score of each item.

\begin{equation}
T_{t_{s},i} = \frac{\sum^{N_{i_{s}}}_{i_{s}=1} \left[\sum^{k_{w}}_{k_{w}=1} s_{c} \left(\frac{t_{vt,c}}{Med(T_{vt})}\right) \frac{m u}{\left(\mu + n_{m}\right)} \right]}{N_{i_{s}}}
\end{equation}

\noindent$T_{t_{s},i}$ - Total trends score for one item\\
$N_{i_{s}}$ - Total number of information sources\\
$i_{s}$ - Source of information\\
$k_{w}$ - Number of keywords in the current item\\
$s_{c}$ - Sentiment score surrounding chosen trend content\\
$m$ - Match value, a Boolean used to check if the current evaluated content contains the chosen trend to be matched against. \\
$u$ - User priority, used to check the current user's interest in the chosen trend. This is 1 by default\\
$t_{vt,c}$ - Tweet volume at this moment in time of the chosen content \\
$Med(T_{vt})$ - Median Tweet volume at this moment in time\\
$\mu$ - Constant, set to 0.1 to avoid division by 0 error for today's trends\\
$n_{m}$ - Number of days between the current day \& the day of the trend.\\


Although the equation supports the calculation of a trend-score using multiple sources of trends, only Twitter trends were used for the testing \& evaluation purposes of this research. 

As much as the constant $\mu$ helps get rid of division by zero error for trends that happened on the same day of calculating the trend-score, it also helps multiply the score by 10 to significantly increase the trend-score of those trends.\\

The beauty of this equation is that it isn't necessarily required to be applied for only \gls{nft} recommendations. It can be used to enhance any content-based recommendations model. It can be seen as another way of infusing collaborative filtering, without the collection of user-specific data by the platform that integrates the presented Recommendations Architecture.

The Total trends score for one item calculated above can either be taken for recommendations as to the top N items or as an absolute similarity match with other chosen items' trends scores. In this research, the author decided to test this with the top N items strategy, to generate featured items recommendations at a given date \& time.

\section{Implementation Choices}

\subsection{Extracting the Keywords of each item}
\gls{nft} asset name, description, collection name \& description were used to extract words that describe the asset. The \textit{RAKE Vectorizer} of the \textbf{NTLK} Python \gls{nlp} library was used for the purpose of extracting keywords from these descriptions.

\subsection{Selection of a Sentiment Analysis Model}
Sentiment scores of the top mentions of the trend on social media (top tweets) were generated using a pre-built sentiment analysis model. 3 Models were tested for this purpose. The 3 models that were tested were:
\begin{enumerate}
\item SpacyTextBlob
\item HappyTransformer
\item Twitter-roBERTa-base for Sentiment Analysis
\end{enumerate}

The 3rd model, which is a state of the art Transformer model \cite{CardiffnlpTwitterrobertabasesentimentHugging, wolfTransformersStateoftheArtNatural2020} that outputs negative, neutral \& positive sentiment scores was chosen. This model outperformed the other two both in the accuracy of the sentiment and the speed. Another advantage of choosing this model was that it was trained on past Tweet data. Therefore, all complexities such as hashtags were handled by the model itself.

This model gave 3 parameters as output. Namely, negative, neutral and positive sentiment scores. The highest score from each of these was taken into consideration.
In the case of negative sentiment, the trend score would become negative. It made sense to leave it as it was since items with negative sentiment may not be suitable to show to users. This could be modified based on the use case of the system.
A neutral sentiment score was taken without any modification as well, while positive sentiment was multiplied by 2. This was done to have a clear increase in trends surrounding positive sentiment since it was expected by the author of the research that a user would be more likely to purchase an item with positive sentiment.

\section{Testing \& Evaluation}

Out of 3872 randomly fetched \gls{nft} asset data \& 677 randomly fetched trends from the OpenSea \& Twitter \gls{api}s respectively, across 14 different datetimes, 55 recommendations were produced using this method. All the trends \& items mined from external APIs were randomly sourced \& heavily pre-processed.

The practicality \& value of the Recommendations Architecture suggested in this research can be better understood by a qualitative evaluation of the outputs produced by the generated graphs that are represented below.

\begin{figure}[h]
     \centering
     \begin{subfigure}[b]{0.47\linewidth}
         \centering
         \includegraphics[width=\linewidth]{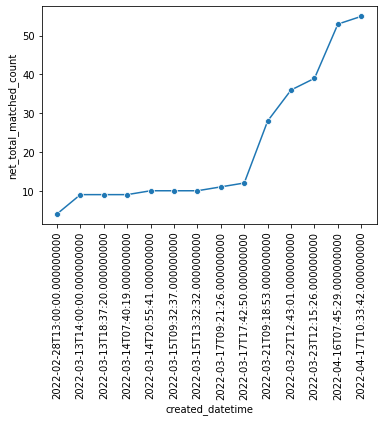}
         \caption{Total Trends based Recommendations made with time}
         \label{fig:trends-recsys-total-matches}
     \end{subfigure}
     \hfill
     \begin{subfigure}[b]{0.47\linewidth}
         \centering
         \includegraphics[width=\linewidth]{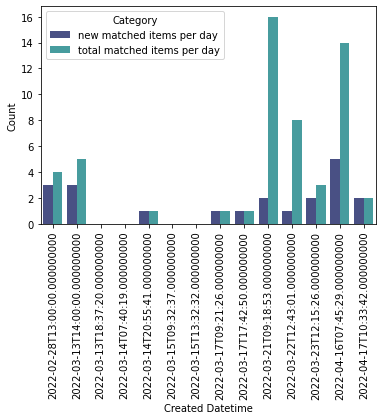}
         \caption{Evaluation of Trends based Recommender}
         \label{fig:trends-recsys-new-matches}
     \end{subfigure}
     \hfill
        \caption{Count of Generated Recommendations Recommendations generated \textit{(self-composed)}}
        \label{fig:counts-of-matches}
\end{figure}

The graph in Figure \ref{fig:trends-recsys-new-matches} shows the count of the items that were matched with the trends of each datetime, highlighting the counts of newly matched items. It is important to note that the rankings of these items that were recommended are updated each time a new set of trends are entered into the system (using an automated process) based on the trend score that is calculated.

The count of matched items in Figure \ref{fig:counts-of-matches} may have been low over the first few days due to fetching worldwide trends, which included trends in languages such as Chinese \& Korean that were not contained in the descriptions of the items. Towards the latter half of the experiment, trends from only the UK were fetched to overcome this constraint.

\begin{figure}[h]
\centering
\includegraphics[width=\linewidth]{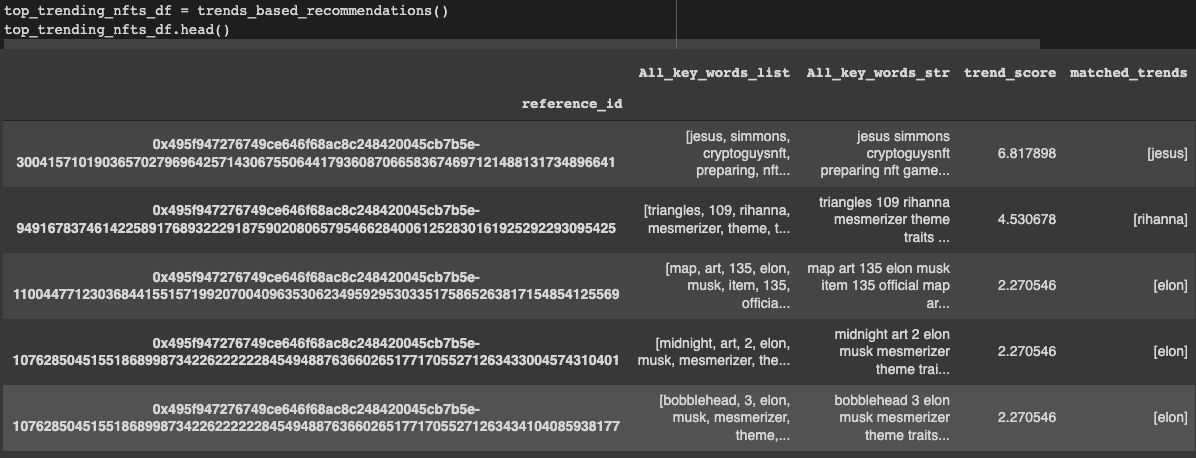}
\caption{Sample output of Generated Recommendations \textit{(self-composed)}}
\label{fig:trends-recsys-sample-output}
\end{figure}

A sample output that was generated by the model from a pandas data-frame output of a Jupyter notebook is shown in Figure \ref{fig:trends-recsys-sample-output}. The keywords have been used to find matches between an \gls{nft} which is referenced by the reference\_id (a combination of the Contract Address of the Smart Contract \cite{IntroductionSmartContracts} that minted the \gls{nft} \& the Token Id of the \gls{nft}). This reference\_id can be used to track the \gls{nft} on the Blockchain.

\begin{figure}[h]
\centering
\includegraphics[width=0.6\linewidth]{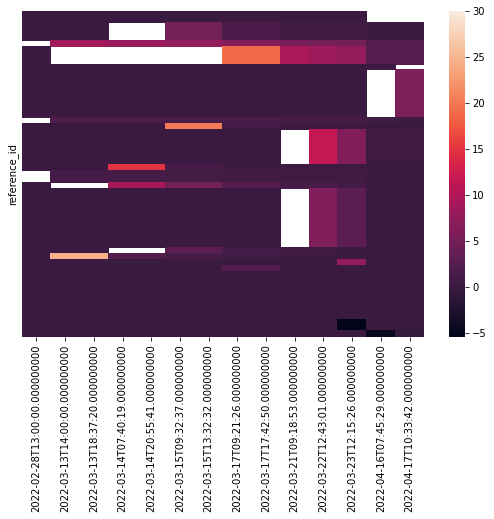}
\caption{Trends based Recommender - Trend Score Heatmap \textit{(self-composed)}}
\label{fig:trends-recsys-heatmap30}
\end{figure}

The heatmap generated by the output produced in Figure \ref{fig:trends-recsys-heatmap30} shows how the trend-score for items decreases with time, from the date of matching with the trend for items that had a maximum trend-score of 30. The max score was limited to generate this heatmap, to make the changes in scores clearly visible for as much items as possible. Additional heatmaps that were generated have been placed in the Appendix of this paper under \textit{\nameref{appendix:extended-testing-eval}}

The matrix that was generated by calculating each trend-score for each \gls{nft} on each datetime of the collected trends can be seen in Figure \ref{fig:trends-recsys-heatmap-data-matrix}.

When taking a look at the heatmap in Figure \ref{fig:trends-recsys-heatmap30}, several observations can be made related to the expected output.
It clearly shows how the trend-score gradually decreases with time, as the trend gets older. The decrease in scores in time can be understood better by the annotated heatmap in Figure \ref{fig:trends-recsys-heatmap-10-anot}.
High-impact trends stay relevant for a longer period of time. Sometimes even better than those matched on the same day. This makes sense since a highly impactful topic is expected to be in peoples' minds for a longer period of time.


Although the trends data matches may be low as it depends on the kind of descriptions of items used for recommendations, another trends recommender could help identify possible interests that a marketplace admin/ creator/ seller could identify for future item additions to an e-commerce platform. This could show the impact of the trend. Even though new matching items could be added over the course of several days, that wouldn't affect the usability of the trend-score calculation algorithm as the number of days since the trend happened is considered for the final trend-score.\\

\section{Conclusion}
In this research, the author had identified the lack of Recommendation Systems for \gls{nft}s. Social media trends data appeared to be a valuable source of getting real-time global trends. An attempt to infuse these trends data into generating valid feature recommendations was attempted.

The trends-based recommender, \textit{\gls{nft}-Trends-RecSys} explored in this research is expected to enhance Content-based Recommendation Systems with Collaborative-filtering-like capabilities, while preserving user anonymity \& without collecting user click-data. The suggested model architecture was tested to identify if it was possible to recommend trending, timely items to users without collecting user-specific information. The results shown are very promising since all the data collected \& used for the research was entirely random \& arbitrary.\\

The data extraction methods explored for recommending \gls{nft}s, integration of social trends into recommendations \& the aggregation algorithm of recommendations utilizing ensembled models are novel results yielded by this research.\\

Many possibilities appeared after the conclusion of this research as mentioned in the Future Work section that could be used as a stepping stone to creating even more interesting \& utilitarian Recommendation Architectures in the future. With the \gls{nft}s expected to be embraced by digital systems \& the internet of the next decade, the outcome of this research \& the invented algorithm could be built-upon to make recommendations as good as those of the last decade.

\section{Future Enhancements \& Novel Possibilities that Emerged}

The current solution does a string match with keywords of each item. This may cause some matches to be skipped due to appearing in different forms. The \gls{nlp} technique, lemmatization could be a possible solution for this. Name Entity Recognition is another \gls{nlp} technique that could enhance the quality of trends data used. The significance of introducing such techniques will have to be tested since they may not have a significant impact on the output as most trends appear to be nouns.

Using multiple sources of trends data would be the first thing the author suggests as this would easily add to the quality and quantity of the generated recommendations. In the case of \gls{nft}s, Reddit \& Discord could be identified as the next 2 best options. Google Trends data \& possibly Search data could be value-addition as well. Furthermore, this can be applied to a localized forum or feedback received in the form of comments on e-commerce sites.

One of the short-comings to help match trends for this purpose that the author of this research noticed is that Twitter trends contain hashtags as trends names at times. Either the developers from the end of Twitter could give a possible solution to it or hashtags may have to be pre-processed and separated.\\

Due to the lack of \gls{nft} data, a \gls{dl} based approach could not be attempted in this research. As a substitute or addition to recently released movies, Amazon's \gls{dl} Neural Network Model could make use of trends, maybe to bolster recommendations for movies as well as e-commerce items.\\

The trends could be categorized to identify similar trends that users seem to show interest in. It would be almost impossible to attempt this level of personalization without collecting user data. Therefore, the value of such an attempt may have to be justified.\\

The trends-based recommender could act as a Decentralized Recommendations System to provide trending recommendations of \gls{nft} assets since the trends and items can come from two different sources. It would be interesting to build a peer-to-peer Recommendation Network that could support this.

The suggested solution to integrate social media trends into recommendations could also help address the cold-start problem in a distributed computing environment or when used as a SASS product where a store gives its items with descriptions and requests for recommendations from a third-party that has social trends data.

\begin{acks}
To Twitter \& OpenSea development teams for providing \gls{api}-keys for free \gls{api} access \& data.
\end{acks}

\printbibliography


\appendix

\section{Extended Testing \& Evaluation of the Model}
\label{appendix:extended-testing-eval}

\begin{figure}[h]
\centering
\includegraphics[width=\linewidth]{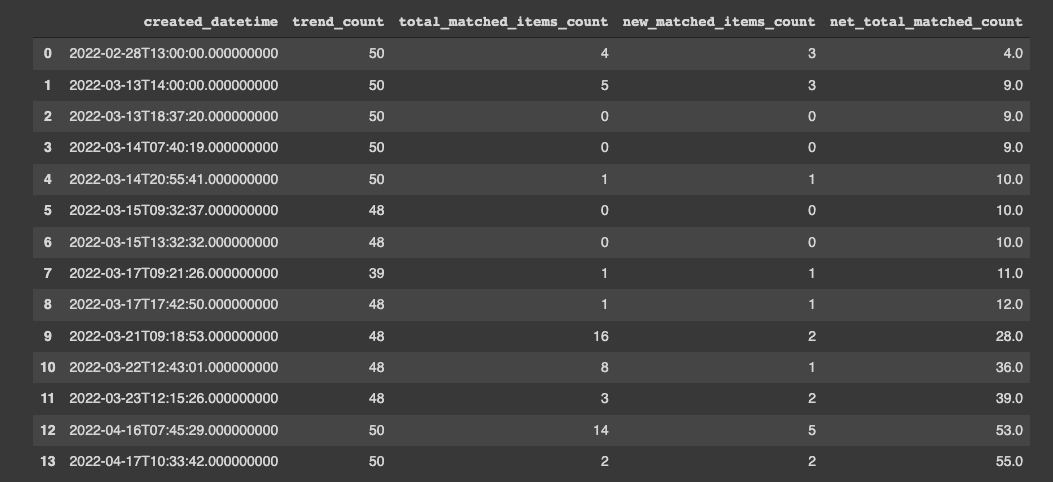}
\caption{Trends based Recommendations Trends Matches Data \textit{(self-composed)}}
\label{fig:trends-recsys-trends-matches-data}
\end{figure}

\begin{figure}[h]
\centering
\includegraphics[width=\linewidth]{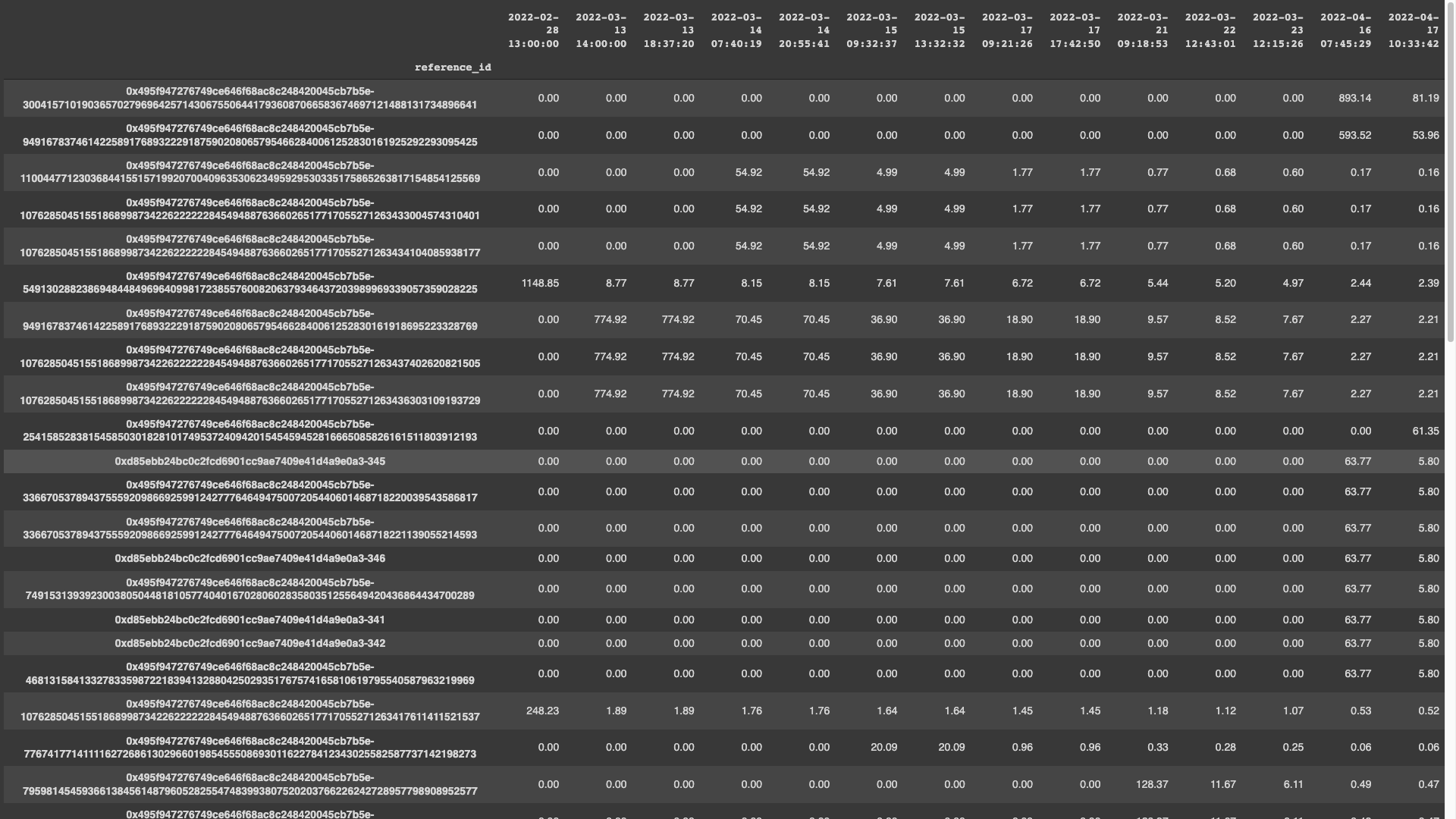}
\caption{Trends based Recommendations Heatmap Data Matrix \textit{(self-composed)}}
\label{fig:trends-recsys-heatmap-data-matrix}
\end{figure}


\begin{figure}[h]
\centering
\includegraphics[width=0.6\linewidth]{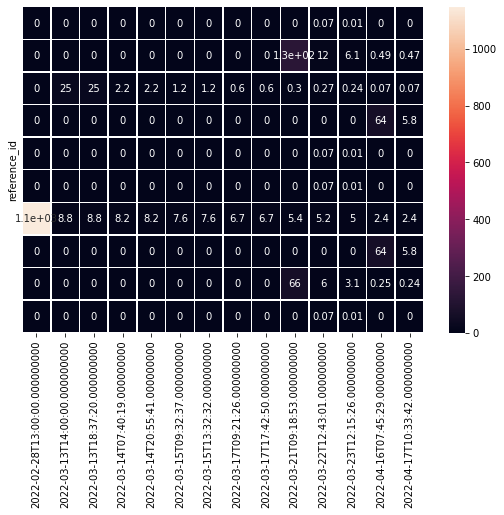}
\caption{Trends based Recommender Testing Annotated Heatmap - 10 random items \textit{(self-composed)}}
\label{fig:trends-recsys-heatmap-10-anot}
\end{figure}

\begin{figure}[h]
\centering
\includegraphics[width=\linewidth]{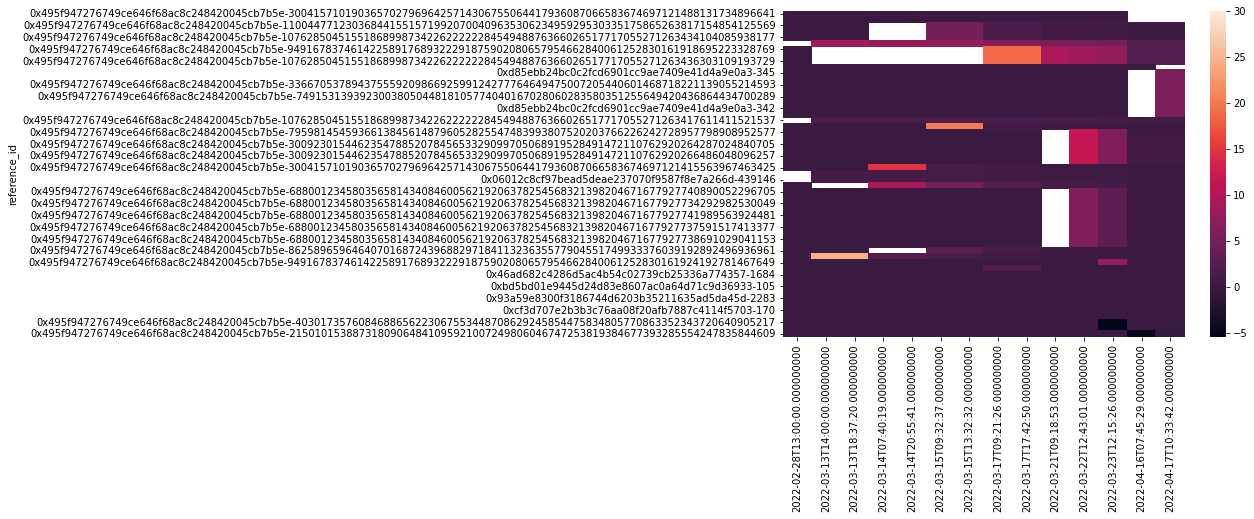}
\caption{Trends based Recommender Testing Heatmap - max score 30 \textit{(self-composed)}}
\label{fig:trends-recsys-heatmap-30-labeled}
\end{figure}

\begin{figure}[h]
\centering
\includegraphics[width=\linewidth]{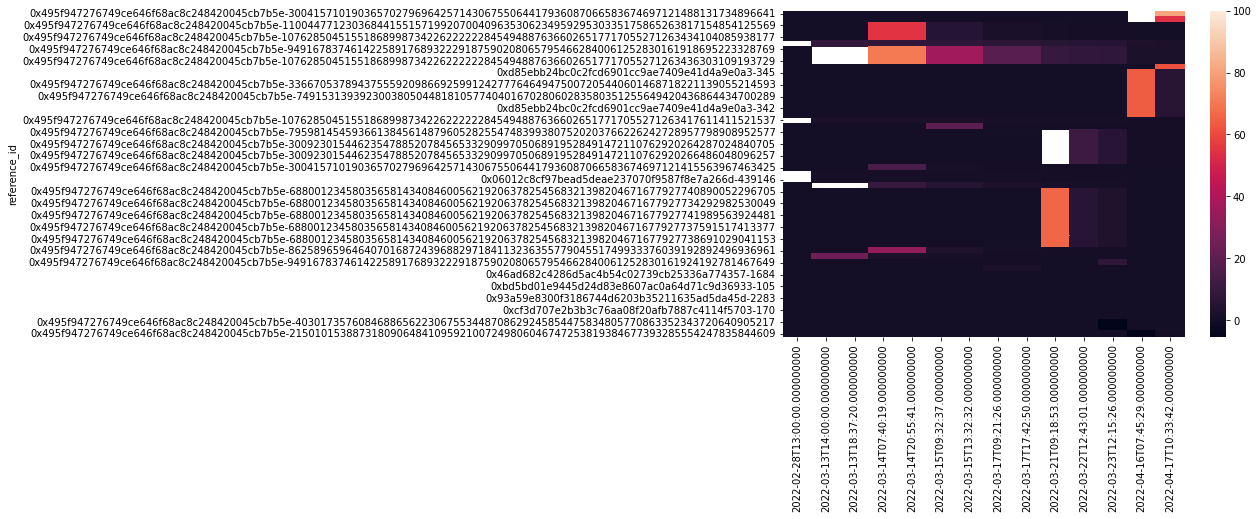}
\caption{Trends based Recommender Testing Heatmap - max score 100 \textit{(self-composed)}}
\label{fig:trends-recsys-heatmap-100}
\end{figure}

\begin{figure}[h]
\centering
\includegraphics[width=\linewidth]{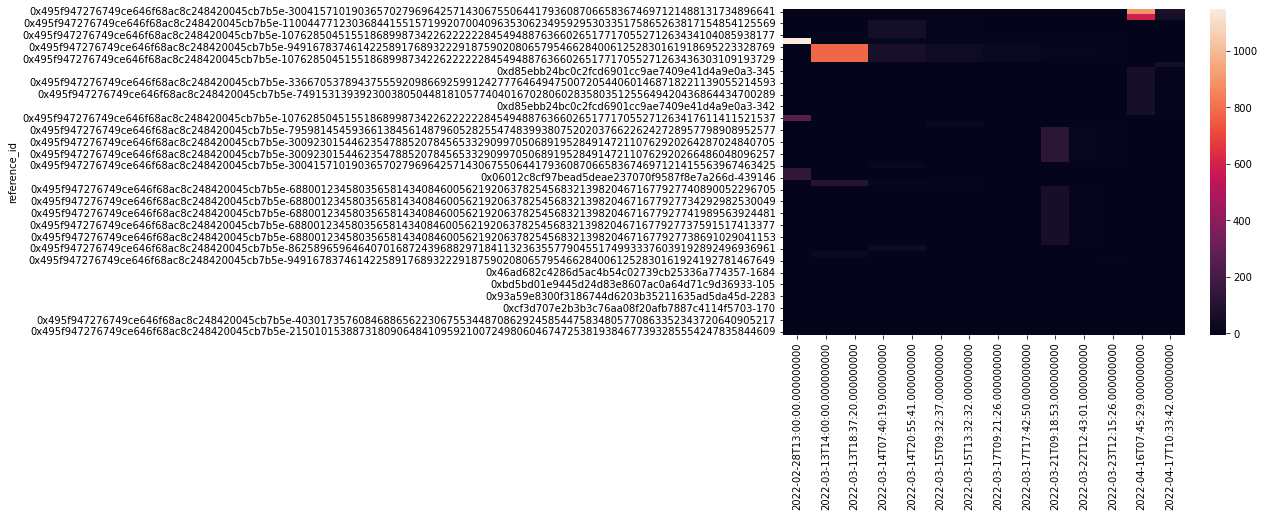}
\caption{Trends based Recommender Testing Heatmap - All items \textit{(self-composed)}}
\label{fig:trends-recsys-heatmap-all}
\end{figure}

\begin{figure}[h]
     \centering
     \begin{subfigure}[b]{0.47\linewidth}
         \centering
         \includegraphics[width=\linewidth]{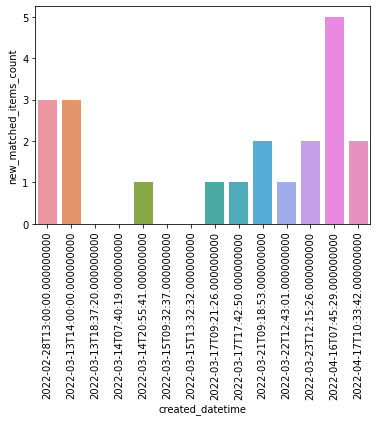}
         \caption{Trends based Recommendations Newly Matched Items}
         \label{fig:trends-recsys-trends-new-matches}
     \end{subfigure}
     \hfill
     \begin{subfigure}[b]{0.47\linewidth}
         \centering
        \includegraphics[width=\linewidth]{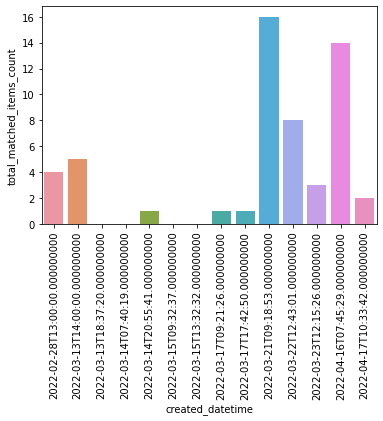}
        \caption{Trends based Recommendations Total Matched Items}
        \label{fig:trends-recsys-trends-total-matches}
     \end{subfigure}
     \hfill
        \caption{Trends based Recommendations Matched Item Counts \textit{(self-composed)}}
        \label{fig:matches-separated-graphs}
\end{figure}

\end{document}